\documentstyle[preprint,aps]{revtex}
\draft
\begin{document}
\preprint{Phys. Rev. Lett. (June 3, 1996) in press.}
\title{Isospin dependence of collective flow in heavy-ion collisions 
at intermediate energies}
\bigskip
\author{\bf Bao-An Li$^{a}$, Zhongzhou Ren$^{b,c}$, C.M. Ko$^{a}$ 
and Sherry J. Yennello$^{d}$}
\address{a Cyclotron Institute and Department of Physics\\ 
Texas A\&M University, College Station, TX 77843, USA\\
b Ganil, BP5027, F14021 Caen Cedex, France\\
c Department of Physics, Nanjing University, Nanjing 210008, P.R. China\\
d Cyclotron Institute and Department of Chemistry\\ 
Texas A\&M University, College Station, TX 77843, USA}
\maketitle

\begin{quote}
Within the framework of an isospin-dependent Boltzmann-Uehling-Uhlenbeck (BUU) 
model using initial proton and neutron densities calculated from the 
nonlinear relativistic mean-field (RMF) theory, we compare the strength of 
transverse collective flow in reactions $^{48}Ca+^{58}Fe$ 
and $^{48}Cr+^{58}Ni$, which have the same mass number
but different neutron/proton ratios. The neutron-rich system 
($^{48}Ca+^{58}Fe$) is found to show significantly stronger negative 
deflection and consequently has a higher balance energy, especially 
in peripheral collisions.
\end{quote}

\newpage
Nuclear collective flow in heavy-ion collisions at intermediate energies 
has been a subject of intensive theoretical and experimental studies 
during the last decade, for a general introduction and overview see
\cite{gary}. The study of the dependence of 
collective flow on entrance channel parameters, such as, the beam energy, 
mass number and impact parameter, have revealed much interesting physics about 
the properties and origin of collective flow. In particular, 
by studying the beam energy dependence it has been found that the 
transverse collective flow changes from negative to positive at an 
energy $E_{bal}$ (defined as the balance energy) due to the competition 
between the attractive nuclear mean field at low densities and the repulsive
nucleon-nucleon collisions\cite{exp1,exp2,exp3,exp4,exp5,exp6,exp7,exp8,exp9}. 
The balance energy was found to depend sensitively on the mass number, impact 
parameter and properties of the colliding nuclei, such as the thickness of 
their surfaces\cite{klakow}. Furthermore, detailed theoretical
studies mainly using transport models (for a review see 
e.g. \cite{sto,bert,bauer92}) have shown that both the strength of transverse 
flow and the balance energy can be used to extract information about 
the nuclear equation of state and in-medium nucleon-nucleon cross sections 
(e.g. \cite{dani85,moli1,moli2,gale87,bert87,tsang,dani89,vd,xu,pan,li93,zhang}).

With high intensity neutron-rich or radioactive beams newly available 
at many facilities, effects of the isospin degree of freedom in nuclear
reactions can now be studied in more detail for a broad range of beam energies 
and projectile-target combinations (e.g.\ \cite{sherry1,gary95}). 
These studies will put stringent constraints on the isospin-dependent part 
of nuclear equation of state. The latter is vital for determining, for example, 
the maximum mass, moment of inertia and chemical composition of neutron 
stars\cite{toki}, where in the crust neutron-rich nuclei coexist with a 
gas of free neutrons and in the core the isospin dependence of the 
nucleon-nucleon interaction determines the stiffness of the equation 
of state\cite{pethick,baym}. In this Letter we report results of the 
first theoretical study on the isospin dependence of transverse flow 
in heavy-ion collisions at intermediate energies. A strong isospin 
dependence of the transverse flow was found at energies around and 
below the balance energy, especially in peripheral collisions. 
An experimental study of the isospin dependence 
of transverse collective flow will soon be carried out at 
NSCL/MSU \cite{gary95}. Detailed comparisons between experimental data and 
model predictions in the future will shed light on the form and strength 
of the isospin-dependent part of nuclear equation of state, the 
isospin-dependent in-medium nucleon-nucleon cross sections, and the 
properties of neutron-rich nuclei.    

In this study we use a Boltzmann-Uehling-Uhlenbeck (BUU) 
transport model which includes explicitly isospin degrees of freedom.
The model has been used recently to explain successfully 
several phenomena in heavy-ion collisions at intermediate 
energies which depend on the isospin of the reaction system\cite{lis95,lir95}. 
The isospin dependence comes into the model 
through both the elementary nucleon-nucleon cross sections $\sigma_{12}$ 
and the nuclear mean field $U$. Here we use the experimental nucleon-nucleon 
cross sections with explicit isospin dependence\cite{data}. We keep in mind, 
however, that in-medium cross sections and their isospin dependence 
might be strongly density dependent\cite{gqli,alm}. 
The nuclear mean field $U$ including the Coulomb and isospin symmetry 
terms is parameterized as
\begin{equation}
      U(\rho,\tau_{z}) = a (\rho/\rho_0) + b (\rho/\rho_0)^{\sigma}\ 
	+(1-\tau_{z})V_{c}+C\frac{\rho_{n}-\rho_{p}}
	{\rho_{0}}\tau_{z}.
\end{equation}
In the above, $\rho_{0}$ is the normal nuclear matter density; 
$\rho$, $\rho_{n}$ and $\rho_{p}$ are the nucleon, neutron and proton 
densities, respectively; $\tau_{z}$ equals 1 or -1 for neutrons or protons, 
respectively; and $V_{c}$ is the Coulomb potential. 
We use the so-called soft equation of state with a
nuclear compressibility of $K=200$ MeV and a value of 32 MeV for 
the strength $C$ of the symmetry potential.

For our first exploratory study on the isospin dependence of the 
collective flow we select two reaction systems, $^{48}Cr+^{58}Ni$ 
and $^{48}Ca+^{58}Fe$, which have the same mass number of $48+58$ but 
different neutron/proton ratios of 1.04 and 1.30, respectively. 
The neutron and proton distributions for these nuclei are determined 
from the well-known relativistic mean-field (RMF) theory 
(e.g. \cite{serot}). More specifically, we use a version of the 
theory with the NL-SH force parameters\cite{sharma1}. The theory describes very 
well the ground state properties of nuclei near and far away from the stability 
line. In particular, it provides a very good description of the neutron-skin 
thickness of nuclei with large neutron excesses\cite{sharma1,sharma2,ren}.  
The neutron, proton and matter densities calculated for the four nuclei studied 
here are shown in Fig.\ 1 and Fig.\ 2. It is seen that 
there are clear neutron skins for the two neutron-rich nuclei $^{48}Ca$ 
and $^{58}Fe$. While the matter densities (n+p) in $^{48}Ca$ and $^{48}Cr$ 
are almost identical, the matter density in $^{58}Fe$ is more 
extended than in $^{58}Ni$. The calculated charge densities of 
these nuclei are actually very close to those measured from electron scattering 
experiments\cite{data2}. In the BUU model, we then initialize the 
spacial coordinates of neutrons and protons in the four nuclei according 
to the calculated densities. The momentum distributions of nucleons are 
generated using the local Thomas-Fermi approximation. It is worth mentioning 
that one can also initialize the neutron and proton distributions by 
running the Vlasov mode of the BUU model for each nucleus. Indeed, 
certain neutron skins can be produced for heavy nuclei by using a strong 
symmetry potentials within the Vlasov model\cite{lis95,lir95,sobotka,joua}. 
Nevertheless, the approach used in the present study is much more reliable 
in terms of reproducing the ground state properties of neutron-rich nuclei.

The standard transverse momentum analysis\cite{dani85} (see also \cite{gary}) 
was performed for the two reaction systems. Typical results for central 
collisions at an impact parameter of 2 fm and beam energies of 50, 60 and 
70 Mev/nucleon are shown in Fig.\ 3. At a beam energy of 50 MeV/nucleon, 
the transverse flow in the reaction of $^{48}Ca+^{58}Fe$ is still negative 
while that in the reaction of $^{48}Cr+^{58}Ni$ is already
positive. The difference disappears at beam energies above 70 MeV/nucleon.
To be more quantitative we have extracted the flow parameter $F$ defined as the
slope of the transverse momentum distribution at the center of
mass rapidity $y_{cm}$. The beam energy dependence of the flow parameter 
for the two reaction systems at impact parameters of 2 fm and 5 fm 
are shown in Fig.\ 4. The lines are the least-square 
fits to the calculations using linear functions $F(Ca+Fe)=-32.2+0.55E/A$ 
and $F(Cr+Ni)=-23.9+0.48E/A$ at b=2 fm; and $F(Ca+Fe)=-35.9+0.22E/A$ 
and $F(Cr+Ni)=-23.2+0.18E/A$ at b=5 fm. It is seen that in 
both central and peripheral collisions the neutron-rich system 
$^{48}Ca+^{58}Fe$ shows systematically smaller 
flow parameters indicating a stronger attractive interaction during the 
reaction. The effect is more appreciable in peripheral collisions as one 
expects. Consequently, the balance energy in $^{48}Ca+^{58}Fe$ reaction
is higher than that in the reaction of $^{48}Cr+^{58}Ni$ by about 10 to 
30 MeV/nucleon. The difference between flow parameters in the two systems 
decreases as the beam energy increases and finally disappears as the 
beam energy becomes far above the balance energy. 

The observed isospin dependence of the collective flow is a result
of the competition among several mechanisms in the reaction dynamics.
First, it is well known that nucleon-nucleon collisions cause repulsive
collective flow, and this effect is proportional to the
number of collisions in the overlapping volume. While the number of particles 
in this volume in the two reaction systems is roughly the same, the number 
of collisions in the reaction of two neutron-rich nuclei is smaller 
since the neutron-neutron cross section is about a factor of three smaller 
than the neutron-proton cross section in the energy region studied here. 
This effect is stronger in peripheral collisions where two thick neutron skins 
are overlapping during the reaction of two neutron-rich nuclei. Second, the 
Coulomb potential also causes repulsive scatterings. This effect is 
obviously weaker in a neutron-rich system. Third, the isospin-independent 
part of the nuclear equation of state is attractive at low densities. 
Since this effect is proportional to the total surface area of the system, 
it increases rapidly with increasing thickness of the colliding
nuclei\cite{klakow}. For neutron-rich nuclei, the nucleon density 
distribution is more extended as shown in Fig.\ 1 and Fig.\ 2. Therefore,
the isospin-independent attractive interaction is stronger in the neutron-rich
system. Finally, the symmetry potential is generally repulsive. One expects a 
stronger effect of the symmetry potential in neutron-rich systems 
in which larger differences between neutron and proton densities exist. 
Although a more quantitative study on the relative importance of these
mechanisms remains to be worked out, it is clear that the 
isospin-independent mean field plays a dominating role in causing the
stronger negative deflection in neutron-rich systems. Moreover, the 
relative effects of these mechanisms depend strongly on the beam energy. 
As the beam energy increases the repulsive nucleon-nucleon collisions 
become dominant and effects of the neutron skin become less important. 
Also, the isospin dependence of the nucleon-nucleon cross sections 
becomes weaker at high energies\cite{data}. It is therefore understandable 
that the isospin dependence of the collective flow disappears at high energies. 

It is well known that the momentum-dependent interaction also affects 
significantly the transverse flow\cite{gale87,zhang,gale90,csernai92,fai}. 
Most importantly, the momentum-dependent interaction gives more weight in terms 
of determining the collective flow to the mean field relative to the 
collision term. The observed stronger negative deflection in the neutron-rich 
system using the momentum-independent equation of state in Eq. 1 would 
therefore be further enhanced by the momentum-dependent interaction.
Consequently, the balance energies in the two systems studied here would be 
even more separated, and this makes the isospin-dependence of the collective
flow to be more easily observable. To quantitatively compare with forthcoming 
experimental data one thus needs to include carefully both the momentum- and 
isospin-dependence of the equation of state in transport models.     

In summary, within the framework of an isospin-dependent BUU model 
using as inputs the neutron and proton density distributions calculated 
from the relativistic mean-field theory, we have demonstrated that 
there is a strong isospin dependence of the transverse collective flow.
The reaction involving neutron-rich nuclei is found to have a significantly 
stronger attractive flow and consequently a higher balance energy 
compared to reaction systems having the same mass number but lower 
neutron/proton ratios. This isospin dependence is mostly easily observed
in peripheral collisions at beam energies around and below the balance energy.  
Our study indicates that the isospin dependence of collective flow may provide
a new approach to extract the isospin-dependent equation of state and to 
investigate properties of neutron-rich nuclei.
 
We would like to thank W. Bauer and G.D. Westfall for their suggestions and 
encouragement to carry out this study. We are also grateful to J.B. Natowitz,
Gongou Xu, Zhongyu Ma and W. Mittig for helpful discussions. This work was 
supported in part by the NSF Grant No. PHY-9212209, PHY-9509266 and 
PHY-9457376, DOE Grant FG05-86ER40256 and the Robert A Welch Foundation 
under Grant A-1266. One of us (ZZR) was supported in part 
by grants from the Foundation of National Educational Commission of 
P.R. China and Ganil in France. One of us (SJY) also acknowledges the support 
from an NSF National Young Investigator Award.

\section*{Figure Captions}

\begin{description}

\item{ Fig.\ 1} \ \ \ 
Proton (dot), neutron (dash) and matter (solid) density 
distributions in $^{48}Ca$ and $^{58}Fe$ calculated using 
the density-dependent relativistic mean-field theory.

\item{ Fig.\ 2} \ \ \
Same as Fig.\ 1 but for $^{48}Cr$ and $^{58}Ni$.

\item{ Fig.\ 3} \ \ \ 
Transverse momentum distributions in the reaction plane as a function of 
rapidity for reactions $^{48}Ca+^{58}Fe$ and $^{48}Cr+^{58}Ni$ 
at an impact parameter of 2 fm and beam energies of 50, 60 and 70 MeV/nucleon.

\item{ Fig.\ 4} \ \ \ 
The flow parameter as a function of beam energy for reactions 
$^{48}Ca+^{58}Fe$ and $^{48}Cr+^{58}Ni$ at impact parameters of 2 
fm and 5 fm. The lines are the least-square fits
to the calculations using linear functions. 
\end{description}
\end{document}